\documentclass{article}
\usepackage[]{alpha}

\usepackage{amsmath}
\usepackage{amssymb}
\usepackage[utf8]{inputenc}
\usepackage{hyperref}

\newcommand{\eq}[1]{eq.~(\ref{#1})}

\newcommand{\fig}[1]{Fig.~\ref{#1}}

\newcommand{\sect}[1]{Sect.~\ref{#1}}

\newcommand{\rmO}{\mathrm{O}}
\newcommand{\rmd}{\mathrm{d}}

\renewcommand{\vec}[1]{\mathbf{#1}}

\def\tr{\mathrm{tr}}
\def\t{\mathrm{T}}
\def\NA{N_\mathrm A}
\def\Nx{N_\mathrm x}
\def\ybar{\bar{y}}
\def\abar{\bar{a}}
\def\dy{\delta \ybar}
\def\da{\delta \abar}
\def\pbar{\bar{p}}
\def\Cw{C_\mathrm{w}}
\def\E{E}
\def\ens{{\cal E}}
\def\NB{N_B}
\def\MC{\mathrm{MC}}
\def\P{\mathcal{P}}

\begin{document}

\title{On fits to correlated and auto-correlated data}
\author[mib,infn]{Mattia Bruno}
\author[desy,hu]{Rainer Sommer}

\address[mib]{Dipartimento di Fisica, Universit\'a di Milano-Bicocca, Piazza della Scienza 3, I-20126 Milano, Italy}
\address[infn]{INFN, Sezione di Milano-Bicocca, Piazza della Scienza 3, I-20126 Milano, Italy}
\address[desy]{Deutsches Elektronen-Synchrotron DESY, Platanenallee~6, 15738~Zeuthen, Germany}
\address[hu]{Institut f\"ur Physik, Humboldt-Universit\"at zu Berlin, Newtonstr. 15, 12489 Berlin, Germany}

\preprintno{DESY-22-154}

\begin{abstract}
    Observables in particle physics and specifically in lattice QCD calculations
    are often extracted from fits.
    Standard $\chi^2$ tests require a reliable determination of the covariance
    matrix and its inverse from correlated and auto-correlated data,
    a challenging task often
    leading to close-to-singular estimates. These motivate modifications of the definition of $\chi^2$ such as uncorrelated fits. 
    We show how the goodness-of-fit measured by their p-value can
    still be estimated robustly for a broad class of such fits.
\end{abstract}

\begin{keyword}
Chi-squared test, Goodness of fit, Autocorrelations
\end{keyword}

\maketitle

\section{Introduction}

In particle physics, physical information is often
extracted from fits to correlated data. When their covariance matrix is known well, 
the minimization of the so-called correlated
$\chi^2$
is the ideal method to obtain the best estimate of
the fit parameters from the statistical point of view.
In many cases model functions are
guided by theoretical principles, but in practice
one has to examine their validity case by case,
and often specific model functions or data points are discarded
based on a measure of their  goodness-of-fit.
For $\Nx$ data points normally distributed, 
and for correlated fits to
model functions depending on $\NA$ parameters,
the expectation
value of $\chi^2$ (at the minimum)
equals the number of degrees
of freedom, $\Nx-\NA$.
When there are many more fitted observables than fit parameters,
a reduced $\chi^2$ close to 1 (defined by the ratio
of the observed $\chi^2$ over $(\Nx-\NA)$), 
provides a first good indication that a given model
describes well the data. 
In addition, the goodness-of-fit  is
judged by computing the probability $Q$, often called p-value, of observing a $\chi^2$
larger or equal than the observed $\chi^2$, which we call $\chi^2_\mathrm{obs}$.
The p-value is easily obtained in closed analytic form 
due to the exact cancellation
between the covariance matrix, $C$, of the underlying data and $C^{-1}$
used in the definition of the correlated $\chi^2$. It depends only on $\chi^2_\mathrm{obs}$ and on
$\Nx-\NA$.

However,  data sets with limited statistics often lead to close-to-singular
$C^{-1}$.
Moreover, for data generated from Monte-Carlo (MC)
methods, specifically from Markov chains, additional complications
arise from the presence of auto-correlations, namely correlations along
the Monte-Carlo history (or ``time'').
Hence for all cases where reliable estimations of the
inverse of the covariance matrix are not possible,
and correlated fits can not be performed,
 the standard $\chi^2$-test is not applicable. 
In this work we point out a robust solution
to the problem.

Our object of study is primarily lattice QCD, where Monte-Carlo
methods based on Markov chains are an essential ingredient
to obtain non-perturbative predictions.
Field configurations are generated by successive repetitions
of a predefined update scheme, and as a consequence,
expectation values of correlation functions, obtained from
averages over measurements on these field
configurations, are both correlated and auto-correlated.
The $\Gamma$-method is the preferable choice to estimate
the statistical errors of primary and derived observables~\cite{Wolff:2003sm}
obtained from a Markov process,
but it is difficult to turn it into a practical method for
general matrix elements of the covariance matrix.\footnote{
A theoretical method is easily given if 1) $\Nx$ is not too large, 2) the exponential 
auto-correlation time, $\tau_\mathrm{exp}$, can be estimated, and 3) the MC chain is very long compared to it. 
Then the window size of Ref.~\cite{Wolff:2003sm} can be chosen a few times $\tau_\mathrm{exp}$ and fixed independently of $i,j$ for 
the estimate of $C_{ij}$ given in the reference. Also jackknife
with sufficiently large bin-size can be used in such, very restricted, cases.}

Because of these issues and the general difficulty in estimating
covariances for large number of data points 
\cite{Michael:1993yj},
ad hoc modifications of $C^{-1}$ are often
used, like so-called SVD cuts 
(see e.g. \cite{Colquhoun:2022atw,Dowdall:2019bea}) or uncorrelated fits. 
We will show how the
expected value of $\chi^2$ and the goodness-of-fit can
be estimated robustly for all these cases. In fact, we emphasize
that this can also be done when the weights of the data points
are independent of the covariance matrix, which may be a good
strategy when fits are used to perform extrapolations away from the region where data
exists.

Our text is organized as follows:
in Section 2 we present the derivation
of the formulae for  the expectation value of $\chi^2$ 
 and the correponding p-value;
in Section 3 we describe how to handle auto-correlations, while
in Section 4 we present a numerical
test in a toy model, before concluding. 

\section{Method}

We want to test whether data $Y_j$ at kinematic coordinates $x_j$
are described by model functions $\Phi(x,a)$,
depending on the parameters $a^\alpha$, 
\begin{equation}
\label{e:model}
Y_j = \Phi(x_j,A) = \phi_j(A)\,,
\end{equation}
with $A^\alpha$ the exact values of
the model parameters: the so-called null-hypothesis. 
In our convention Greek indices run from $1$ to $\NA$,
Roman ones run from $1$ to $\Nx$, repeated indices are summed over, and we use the natural scalar product $(r,s)=r_is_i$ and norm $||r||=(r,r)^{1/2}$.
We also often use the matrix-vector 
notation to suppress Roman indices.
Given data $\ybar_j$ with normal distribution,
\begin{equation}
P_C(\ybar) = (2\pi)^{-\Nx/2} (\det C)^{-1/2} \exp \Big( -\frac12 (\dy, C^{-1} \dy) \Big) \,, \quad
\, \quad \dy = \ybar -Y \,,
\label{e:gauss}
\end{equation}
with covariance matrix $C$,
we want to find robust estimates for the fit parameters
and have a measure for the likelihood that our model
is compatible with the data. Expectation values
are defined in the usual manner. In particular we have
\begin{eqnarray}
    \langle \dy_i \dy_j \rangle \equiv \int \rmd \ybar \
    P_C(\ybar) \ \dy_i \dy_j = C_{ij} \,.
\end{eqnarray}
One possibility when such a normal distribution is approximately realized is when
$\ybar_j$ are averages over Monte Carlo data of length $N$
and $N$ is sufficiently large. In this case,
all elements of $C$ are of order $1/N$, and $\langle \ybar_i \rangle \to Y_i$ in the
limit $N \to \infty$.

The natural measure for the distance of the model to the data is
\begin{equation}
\chi^2(a) = ||W (\ybar-\phi(a))||^2  \,,
\end{equation}
in terms of a symmetric and positive $\Nx \times \Nx$ weight matrix $W_{ij}$.
A reasonable requirement is that $\chi^2$ remains finite
as $N\to\infty$, which means that $W$ should be of $\rmO(N^{1/2})$:
common choices for $W^2$ are the inverse covariance matrix $C^{-1}$
or the inverse of its diagonal projection 
$W_{ij}=\delta_{ij} / \sqrt{C_{ii}}$ (no summation over $i$).
The latter definition of $\chi^2$ leads to so-called uncorrelated fits. However, one can also consider weight matrices $W$ which 
scale weaker with $N$, e.g. $\rmO(N^0)$. We will remark on that later.

As usual, from the minimization of the $\chi^2$ function 
we obtain
the best estimate of the parameters
\begin{equation}
\chi^2(\abar) = \underset{a}{\mathrm{Min}}\, \chi^2(a)  \,, \quad
\abar = \abar(\ybar) \,.
\label{e:minimum}
\end{equation}
Here we assume that the position $\abar(\ybar)$ of the minimum of $\chi^2$ is unique.
Our goal is to derive an expression for the expectation value
\begin{equation}
  \langle \chi^2(\abar)\rangle =
  \int \rmd\ybar \ P_C(\ybar) \ \chi^2(\bar{a}(\ybar)) \,,
\label{e:chiexp0}
\end{equation}
predicted by the normal distribution of the data and under the assumption that Eq.~\eqref{e:model} holds.

The  condition for $\abar$, 
\begin{equation}
	(W\phi^\alpha(\abar),W(\ybar -\phi(\abar)) =0 \,, \quad
 \text{with } \phi^{\alpha_1 \cdots \alpha_m}(a) \equiv \frac{\partial^m \phi(a)}{\partial a^{\alpha_1} \cdots \partial a^{\alpha_m}} \,,
\end{equation}
can be rewritten as 
\begin{equation}
	\P\, W\, (\ybar -\phi(\abar)) =0 \,,
	\label{e:minimum2}
\end{equation}
with $\P$ the projector onto the space spanned by $\left\{W\,\phi^\alpha\right\}$, namely
\begin{equation}
  \P W\phi^\alpha = W\phi^\alpha\,, \; \P = \P^2\,,\; \tr \P = \NA \,, 	
\end{equation}
and we have
\begin{equation}
    \chi^2(\abar) = ||(1-\P)W\,(\ybar-\phi(\abar) ) ||^2 \,.
\end{equation}

\subsection{Expectation value of $\chi^2$}

We expand $\phi$ around $a=\abar$, 
\begin{equation}
    \phi(A) = \phi(\abar) -  \phi^\alpha(\abar) \da^\alpha + 
    \frac12 \phi^{\alpha\beta}(\abar) \da^\alpha\da^\beta + \rmO(\da^3)\,,
    \quad \da^\alpha \equiv \abar^\alpha - A^\alpha \,, 
\end{equation}
and obtain, using $(1-\P) W \phi^\alpha(\abar)=0$,
\begin{equation}
    \chi^2(\abar) = ||(1-\P)W\, \dy  ||^2 +
    (W\dy,(1-\P)W \phi^{\alpha\beta}) \da^\alpha\da^\beta +
    \rmO( N^{-1})\,.
    \label{e:minimum3}
\end{equation}
Taking $\da=\rmO(N^{-1/2})=\dy$, the expected value of $\chi^2$ 
is then
\begin{equation}
  \langle \chi^2(\bar a) \rangle = \tr [\Cw  (1  -  \P) ] +\rmO(N^{-1})\,, \quad \Cw = W C W \,.
\label{e:chiexp1}
\end{equation}
We stress the fact that in the result above even an approximate knowledge 
(up to statistical errors) of the covariance matrix is sufficient.
The corrections to \eq{e:chiexp1} are seen to be $\rmO(N^{-1})$ in the following way.
Expanding $\da(\dy)$ in terms of $\dy$ (and therefore in terms of $N^{-1/2}$), 
we have $\langle (W\dy,(1-\P)W \phi^{\alpha\beta}) \da^\alpha \da^\beta \rangle = g_{ijk} \langle \dy_i\dy_j\dy_k\rangle +\rmO(N^{-1})$ 
where $g=\rmO(N)$ due to the two factors of $W$.
In the assumed symmetric (normal) distribution $\langle \dy_i\dy_j\dy_k\rangle$ vanishes and skewed corrections to it are down by a factor $N^{-1/2}$
leading overall to $\rmO(N^{-1})$.
However, in practice the estimate of $\Cw$ from the data 
themselves introduces order $N^{-1/2}$ uncertainties, as well as the 
projector $\P$ which is a function of $\abar^\alpha$.
The essential point is that $\langle \chi^2(\abar) \rangle$ can 
easily be computed also with auto-correlated data, as we will show in the next section.

For the special choice of an ``uncorrelated fit'', i.e. 
 $W_{ij}=\delta_{ij} / \sqrt{C_{ii}}$, 
we have $[\Cw]_{ii}=1$ and
\begin{equation}
 \langle \chi^2(\abar) \rangle = \Nx -\tr [\Cw  \P ] \,.
\label{e:chiexp2}
\end{equation}
Similarly, for a ``correlated fit'' with $W^2=C^{-1}$,
$[\Cw]_{ij} = \delta_{ij}$,   
we reproduce the standard result $\langle \chi^2(\abar) \rangle = \Nx-\NA$.

Above we used $W = \rmO(N^{1/2})$.  However, one could also consider a more general 
functional dependence of $W$ on $C$, for instance by taking 
$W$ independent of $C$\footnote{Consider for example a linear extrapolation of 
$\Phi(x,a) = a_0 + a_1 x +\rmO(x^2)$, to $x=0$, i.e. we want to determine $a_0$. When the data set is not 
rich enough to include higher order terms in $\phi$, it can be 
a good strategy to choose a weight $W_{ii} = 1/x_i^n$, e.g. with $n=2$. 
In this case $W$ is $\rmO(N^0)$.}. 
This changes nothing in the above, 
since $\da$ remains $\rmO(N^{-1/2})$.
In Appendix~\ref{a:generalizations} we discuss several other 
generalizations of the fit functions assumed above.

At this point, 
it is natural to define the reduced-$\chi^2$ for fits with arbitrary $W$ as 
$\chi^2/\langle \chi^2(\abar) \rangle$ with 
eq.~(\ref{e:chiexp1}) for $\langle \chi^2(\abar) \rangle$.
Although a reduced-$\chi^2$ of order 1 is
already a first indication of whether a given model fits well the data,
the real discriminator for the goodness of a fit is the p-value.

\subsection{Quality-of-fit}

The probability for having  a value
of $\chi^2(\abar)$ which is larger or equal than 
a given value $\chi^2_\mathrm{obs}$ is
\begin{equation}
    Q = \int \rmd \ybar \, P_C(\ybar) \,
    \theta\big(\chi^2(\abar(\ybar)) - \chi^2_\mathrm{obs} \big) \,.
\end{equation}
It is useful to decide whether the model can be statistically rejected
(the null hypothesis).
In practice $\chi^2_\mathrm{obs}$ stands for the value of
$\chi^2(\abar)$ that one has in the particular fit
to the existing data and $Q$ gives the probability of finding a fit worse than the one that one found, i.e. with a larger value of $\chi^2(\abar)$. 
If $Q$ is too low,
then one has
significant reasons to doubt that the model
describes the data, and usually dismisses the fit
and modifies the model or the fit ranges. The probability $Q$ is often called quality-of-fit or p-value.

Using the variables $z=C^{-1/2} \dy$
it is given by
\begin{equation}
   Q(\chi^2_\mathrm{obs},\nu) = 
    \int \rmd z \, (2\pi)^{-\Nx/2} e^{-\frac12 ||z||^2}
            \; \theta \big( (z,\nu z) -\chi^2_\mathrm{obs} \big)\,,
\end{equation}
in terms of the positive, symmetric matrix
\begin{equation}
     \nu = C^{1/2} W (1  - \P)  W C^{1/2}\,.
     \label{e:nu}
\end{equation}
The latter takes over the role of the number of degrees of freedom
in correlated fits, where $CW=1$ and $\tr[\nu]=\Nx - \NA$.
Denoting the strictly positive eigenvalues of $\nu$ by
$\lambda_i(\nu)\,,i=1\ldots N_\nu \leq \Nx-\NA$,
we have
\begin{equation}
    Q(\chi^2_\mathrm{obs} , \nu)
           = \int \theta\big( 
            \sum_{j=1}^{N_\nu}\lambda_j(\nu) z_j^2- \chi^2_\mathrm{obs} \big) \; \prod_{i=1}^{N_\nu} \frac{\rmd z_i}{ (2\pi)^{1/2}}\;
            e^{ -\frac12 z_i^2}
            \,\,.
            \label{e:Qint}
\end{equation}
In this form, $Q$
is easily evaluated by a MC over $z_i$ drawn from a Gaussian distribution. A sample of order thousand random numbers is sufficient for a precision
of $Q$ of order 0.01, and our proposal is to
use this as estimate for the quality-of-fit.

The difficult part is not the statistical error in
the MC estimate of \eq{e:Qint} but the fact that
$\nu$ has to be estimated from the data $\bar y$ and thus the eigenvalues
$\lambda_i(\nu)$ are uncertain themselves. This leads to the expectation that the statistical error $\Delta Q$ of $Q$ is order $N^{-1/2}$.
In contrast to the case of a correlated $\chi^2$ 
fit no enhancement by small eigenvalues (of the estimated $\nu$) will occur in $\Delta Q$. 
Instead the contribution of modes  with small eigenvalues are suppressed. In sect.~\ref{s:toy} we will demonstrate 
in a toy model that $\Delta Q$ is small and can be neglected in practice. 

In general one should decide whether a fit 
is to be discarded based on the quality-of-fit. For example,
if $Q(\chi^2_\mathrm{obs} , \nu)=0.05$ there is only a 5\% chance that 
one finds such a $\chi^2_\mathrm{obs}$ or worse but the fit-function describes the data on average. One will then 
dismiss fits with $Q(\chi^2_\mathrm{obs} , \nu)\leq 0.05$
or with a different threshold.
When one performs a correlated fit and has a large number of data, more precisely when
$\Nx-\NA=\langle \chi^2(\abar) \rangle\gg1$, then
$Q(\chi^2_\mathrm{obs} , \nu)$ 
is close to a step function $1-\theta(\chi_\mathrm{obs}^2/\langle \chi^2(\abar) \rangle)$. But this is not the case in general. 
To understand this behavior, we examine the region where
$Q$ makes the transition from $Q\approx1$ to $Q\approx0$. Its width  can be 
estimated by the variance 
of $\chi^2(\abar)$. Using eqs.~(\ref{e:minimum3},~\ref{e:chiexp1}) we obtain
\begin{equation}
  \langle [\chi^2(\abar) - \langle \chi^2(\abar) \rangle]^2 \rangle =
    \langle (\chi^2(\abar))^2 \rangle - \langle \chi^2(\abar) \rangle^2 =
    2 \tr(\nu^2) + \rmO(N^{-1/2}) \,,
\end{equation}
valid for arbitrary weights $W$.
For a correlated fit with $WCW=1$, the width is suppressed by $1/(\Nx-\NA)$
relative to $\langle \chi^2(\abar) \rangle^2$.  
For a large number of data points,
a step-function behavior is approached. However, in general this is
not the case. Assume for example that the eigenvalues of $\nu$ are 
geometrically distributed, namely $\lambda_j=\lambda_0 r^j$ with $r<1$. Then we have
\begin{equation}
    \frac{\langle [\chi^2(\abar) - \langle \chi^2(\abar) \rangle]^2 \rangle}{2\, \langle \chi^2(\abar) \rangle^2}	
    = \frac{\sum_{j=1}^{N_\nu}(\lambda_j(\nu))^2}{(\sum_{j=1}^{N_\nu}\lambda_j(\nu))^2} = 
    \frac{1-r}{1+r} +\rmO(r^{N_\nu+1}) \,.
\end{equation}
In general the width is therefore not suppressed at all and it is necessary to examine the p-value.
For an illustration the reader may look ahead at
$Q$-functions in our toy model in \fig{f:pvalue}.

\section{Auto-correlated Monte Carlo data}
\label{s:autocorr}

In the above formulae the covariance matrix is assumed to be known,
but in practice we have to replace it
by an estimate from the Monte Carlo data, often in presence
of non-negligible auto-correlations, which we discuss next.

We consider explicitly the case where MC data from Markov chains with different update
schemes or different equilibrium distributions enter the fit.
The specialization to multiple MC chains with the 
same update scheme, often called replica, is straightforward.
On a single MC chain labelled by $\ens$ we assume to have estimators  
$p_i^{\ens}(t)$ at MC time $t$ of several quantities labelled by $i$, such that 
\begin{equation}
    \pbar_i^\ens = \frac{1}{N_\ens} \sum_{t=1}^{N_\ens} p_i^\ens(t)
\end{equation}
converges to $P_i$ for $N_\ens\to\infty$.   In general, the $Y$-variables in our fits are functions
\begin{equation}
 Y_j=\eta_j(P)
\end{equation}
of primary observables $P$, estimated by $\ybar_j=\eta_j(\pbar)$.
Note that if different chains yield estimators for 
the same observable $P_i$, we just replace 
$P_i$ by 
the (weighted) average over the different chains. This average is then considered as part of the function $\eta_j$.

\subsection{$\Gamma$-method}

The preferred method for dealing with auto-correlations
in the error estimate of $\pbar_i^\ens $ is the so-called $\Gamma$-method \cite{Madras:1988ei,Wolff:2003sm}.
Different error estimators are discussed in Appendix~\ref{a:errors}.
We refer the interested reader to Ref.~\cite{Kelly:2019wfj} 
for a proposal to estimate $\langle \chi^2(\abar) \rangle$ and p-value 
based on bootstrap resampling, 
from data which are blocked in MC time with a block-length 
large compared to the exponential auto-correlation time 
(see also Refs.~\cite{RBC:2020kdj,RBC:2021acc} for practical applications).

The $\Gamma$-method inherits its name from the matrix valued auto-correlation function
\begin{equation}
    \Gamma^\ens_{ij}(t) =   \langle  \Delta p^\ens_i(t+t_0) \,\Delta p^\ens_j(t_0)\rangle_\ens \,, \quad \Delta p^\ens_i(t)=p^\ens_i(t)-P_i\,.
\end{equation}
Here the average $\langle.\rangle_\ens$ may be defined 
as the $t_0$-average of an infinitely long chain. 
Linearizing the error propagation and
applying the chain rule as in \cite{Wolff:2003sm}
leads to the same simple form above ($\Delta p \to \Delta y$) also for the auto-correlation
function of derived quantities, with
the linearized fluctuations given by
\begin{equation}
    \Delta y_i^{\ens}(t) = \sum_k \frac{\partial \eta_i}{\partial p_k} 
    \bigg\vert_{p=P} \Delta p^\ens_k(t) \,.
\end{equation}
Since
the covariance matrix of the variables $\ybar_i^\ens$ reads
\begin{equation}
    C_{ij}^\ens =  \frac{1}{N_\ens} \sum_{t=-\infty}^\infty \Gamma_{ij}^\ens(t) \,,
    \label{e:covmatac}
\end{equation}
the expected $\chi^2$, 
\eq{e:chiexp1}, becomes
\begin{equation}
    \langle \chi^2(\abar) \rangle = \sum_\ens \frac{1}{N_\ens} \sum_{t=-\infty}^\infty
    \tr \big[ \Gamma^\ens(t) \, W (1  -  \P) W\big] \,,
    \label{e:chiexpac}
\end{equation}
in presence of auto-correlations. 
With finite length MC data
the above formulae need to be supplemented
by summation windows in $t$, over which
the MC estimate of $\Gamma_{ij}^\ens(t)$,
denoted by $\Gamma_{ij}^{\ens,\MC}(t)$, 
should be summed \cite{Madras:1988ei,Wolff:2003sm,Schaefer:2010hu}.
Hence a good approximation to \eq{e:chiexpac} is found in the form
\begin{eqnarray}
    \langle \chi^2(\abar) \rangle
    &=&E_f \pm \Delta E_f\,,
    \quad E_f = \sum_\ens E_f^\ens(w_\ens) \,,
    \label{e:chiexp-ef}
    \\
    E_f^\ens(w_\ens)&=& \frac{1}{N_\ens} \sum_{t=-w_\ens}^{w_\ens}
    \tr [\Gamma^{\ens,\MC}(t) \, W (1  -  \P) W] \,,
    \label{e:res1}
\end{eqnarray}
or, assuming $W^{-2}=\mathrm{diag}(C)$ (uncorrelated fit),
\begin{eqnarray}
    \langle \chi^2(\abar) \rangle
    &=&\Nx-E_s \pm \Delta E_s\,,\quad
    E_s = \sum_\ens E_s^\ens(w_\ens) \,,
    \label{e:chiexp-es}
    \\
    E_s^\ens(w_\ens)&=& \frac{1}{N_\ens} \sum_{t=-w_\ens}^{w_\ens}
    \tr [\Gamma^{\ens,\MC}(t) \,  W\P W]\,.
    \label{e:res2}
\end{eqnarray}
In both cases the standard automatic windowing procedure of
Ref.~\cite{Wolff:2003sm} or the one of Ref.~\cite{Schaefer:2010hu}
may be applied simply replacing $\Gamma_F(t)$ in Ref.~\cite{Wolff:2003sm}
by $\tr [\Gamma(t) \, W \P W]$ or
$\tr [\Gamma(t) \, W (1-\P) W]$,
while nothing changes in the discussions in Refs.~\cite{Wolff:2003sm,Schaefer:2010hu}.
Note that the knowledge of the full covariance matrix is not required.
Its presence is captured by the traces in \eq{e:res1} and \eq{e:res2}.
We may use the same automatic determination of the windows ensemble by ensemble and
possibly the addition of a tail~\cite{Schaefer:2010hu} due to large auto-correlations
and will have balanced statistical and systematic errors
(from the truncation of the $t$-sum) of size
\begin{equation}
    \Delta E_{f,s}^2 \approx 2
    \sum_\ens \frac{2w_\ens+1}{N_\ens} \, [E_{f,s}^\ens(w_\ens)]^2\,.
    \label{e:errEfs}
\end{equation}
Assuming that the required $w_\ens$ are not much different,
for strong correlations (not auto-correlations)
\eq{e:res1} will be more precise, while in
the opposite case \eq{e:res2} will be better.
Obviously one may just evaluate
both and take the one with the smaller error.

The covariance matrix may be estimated
from $\Gamma_{ij}^{\ens,\mathrm{MC}}$
by truncating the $t$-sum in \eq{e:covmatac}
to the summation windows $w_\ens$ obtained from
the automatic procedure for $E_{f,s}^\ens$.
The  p-value is 
then obtained from \eq{e:Qint}. In \sect{s:toy} we demonstrate in a toy model that this is a valid approach, namely 
the p-value is stable against increasing the window size. Note that only modes with 
$\lambda(\nu)>0$ enter \eq{e:Qint}. 

Two remarks on the error estimates, \eq{e:errEfs} are in order. These are of the type error-of-the-error. 
As is common\cite{Madras:1988ei,Wolff:2003sm,Schaefer:2010hu}, we use the Madras-Sokal approximation, which consists of neglecting the connected parts of a dynamical four-point function. We are not aware of
a general argument on the quality of this approximation, but F.~Virotta found that it 
works quite well in a simple interacting field theory \cite{Virotta2012Critical}.
Secondly \eq{e:errEfs} neglects the statistical error of $\P$ which
is a function of $\abar$ and of the same order in $N^{-1}$ as $\Delta E_{f,s}$. In Appendix~\ref{a:proof} we provide the analytic formula for
its evaluation and in all our experiments we find it to be subdominant compared to 
the terms in \eq{e:errEfs}.

Note that when the data can be partitioned in sufficiently many  and large blocks,
i.e. when autocorrelations are not an issue, one may consider the bootstrap
approach to get an estimate of the error of the p-value.

\section{Numerical illustration}
\label{s:toy}

\subsection{Toy model}

As an illustration, we consider the free one-component
scalar theory with standard action,
\begin{equation}
  S = a^d \sum_x \, \bigg[ \frac12 \partial_\mu \phi(x)\partial_\mu \phi(x)
    + {m^2\over 2} \phi^2(x)  \bigg]
\end{equation}
on a $T\times L^d$ torus
(periodic boundary conditions) discretized on a hyper-cubic
lattice with spacing $a$ and standard discrete forward derivative $\partial_\mu$.
Our observables are the often used time-momentum correlation
functions
\begin{equation}
    G_{\phi\phi}(t,\vec p) =
    a^{d} \sum_\vec{x} e^{-i\vec p \cdot \vec x}\langle \phi(t, \vec x) \phi(0) \rangle \,,
    \quad
    G_{\phi^2\phi^2}(t)=
    a^{d} \sum_\vec{x} \langle \phi^2(t, \vec x) \phi^2(0) \rangle \,. 
\end{equation}
In absence of interactions, the Green function $G_{\phi\phi}$
on a torus with finite time extent $T$ is given by 
\begin{equation}
    G_{\phi\phi} (t,\vec{p}) = \sum_{n=-\infty}^{\infty} \Delta(t +n T,\vec p)\,,
    \label{e:Gphiphi}
\end{equation}
in terms of the infinite $T$  propagator
\begin{equation}
    \Delta(t,\vec p) = \frac{e^{-|t|\,\omega(\vec p)}}{2
    \sinh(a\omega(\vec p))/a} \,,
\end{equation}
where $\omega(\vec p)$ satisfies the lattice dispersion relation
\begin{equation}
  2[\cosh(a \omega(\vec p))-1] = a^2 m^2 + 2\sum_{i=1}^d (1-\cos(ap_i))\,,
\end{equation}
and $p_i=\frac{2\pi}{L} k_i\,, k_i=0,1, \ldots L/a-1$.
A further correlator is
\begin{equation}
    G_{\phi^2\phi^2} (t) = \frac{2}{L^d} \sum_{\vec p} G_{\phi\phi}(t,\vec p)^2 +
    \frac{1}{L^d} \Big[\sum_{\vec p} G_{\phi\phi}(0,\vec p)\Big]^2 \,,
    \label{e:Gphi2phi2}
\end{equation}
where the second term is a disconnected contribution
which dominates at long distances.

On a given MC configuration, one may estimate $G_{\phi\phi}$ from 
\begin{equation}
    \E_{\phi\phi} (t,\vec p) =
     \tilde \phi(t,\vec p) \,\tilde\phi(0,-\vec p)\,,
    \quad \tilde\phi(t,\vec p)=\frac{a^{d}}{L^{d/2}}
    \sum_\vec{z} e^{-i\vec p \cdot \vec z} \phi(z_0, \vec z) \,,
    \, \label{e:c2}
\end{equation}
and for the case of negligible auto-correlations,
i.e. measurements separated by MC time much larger than
the exponential  auto-correlation time,
the covariance matrix of $N$ measurements of the estimators $\E(t,\vec p)$
takes the simple form ($g(t)\equiv G_{\phi\phi}(t,\vec p)$)
\begin{equation}
  N \times \mathrm{Cov}(t,t')  =
  g(t) g(t')
  +\delta_{\vec p,0}\,g(t-t') g(0) \,.
  \label{e:Gphiphi-cov}
\end{equation}
We note that off-diagonal matrix elements are exponentially suppressed
by $e^{-|t-t'|m}$ and there is the typical strong exponential
noise/signal enhancement at large $t \approx t'$. 
For the noisy $G_{\phi^2\phi^2}$ correlator,
we employ an estimator with a maximal 
volume average, 
\begin{equation}
    \E_{\phi^2\phi^2} (t) =
    \frac{a^{d+1}}{T L^d} \sum_{z_0=0}^{T-a} \overline{ \phi^2}(z_0+t) \,
    \overline{ \phi^2}(z_0) \,,
    \quad \overline{ \phi^2}(z_0) \equiv
    \sum_\vec{z} \phi^2(z_0, \vec z) \,,
    \, \label{e:c2}
\end{equation}

\subsection{Simulation}

To illustrate our method
on auto-correlated data similar to standard lattice QCD calculations,
we simulated the free theory in 1+1 dimensions 
with the following algorithm\footnote{See Refs.~\cite{Bruno:2016plf,Bahr:2019eom} for 
applications of our method to uncorrelated fits performed on real 
Lattice QCD data.}. We performed a MC iteration composed of a Hybrid Monte Carlo (HMC)
trajectory of length  4 molecular dynamics 
units \cite{Duane:1987de}  followed by one Metropolis
sweep. Both the integration of the equations
of motion in the HMC and the width of the Metropolis proposal were tuned to have $\approx80\%$ acceptance. The large trajectory length was chosen 
to tame auto-correlations and the Metropolis sweep was added to 
avoid exceedingly large auto-correlations, specific to the free theory \cite{Kennedy:2000ju}. 
\begin{figure}[hb]
    \centering
    \includegraphics[width=.49\textwidth]{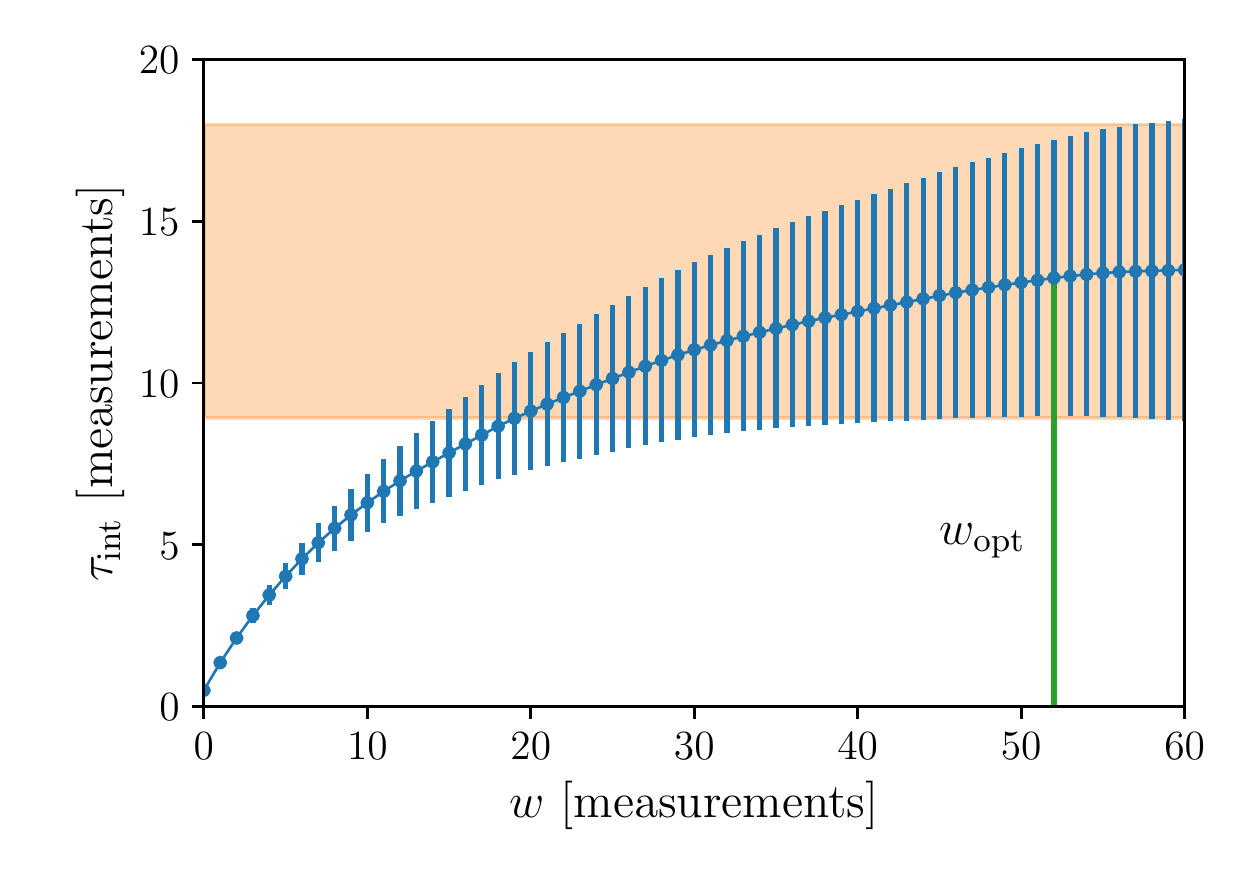}
    \caption{
    Determination of the integrated auto-correlation time, $\tau_\mathrm{int}$, 
    of $\langle \phi^2 \rangle$ with the window size $w_\mathrm{opt}$ determined by 
    $\tau_\mathrm{int}=4 w_\mathrm{opt}$. 
    }
    \label{f:autocorrphiphi}
\end{figure}
Setting $am=0.075$, we simulated a lattice of size $160 \times 40$ 
and generated several replica of 2000 consecutive configurations 
with the algorithm described above.
\begin{figure}[ht]
    \centering
    \includegraphics[width=.49\textwidth]{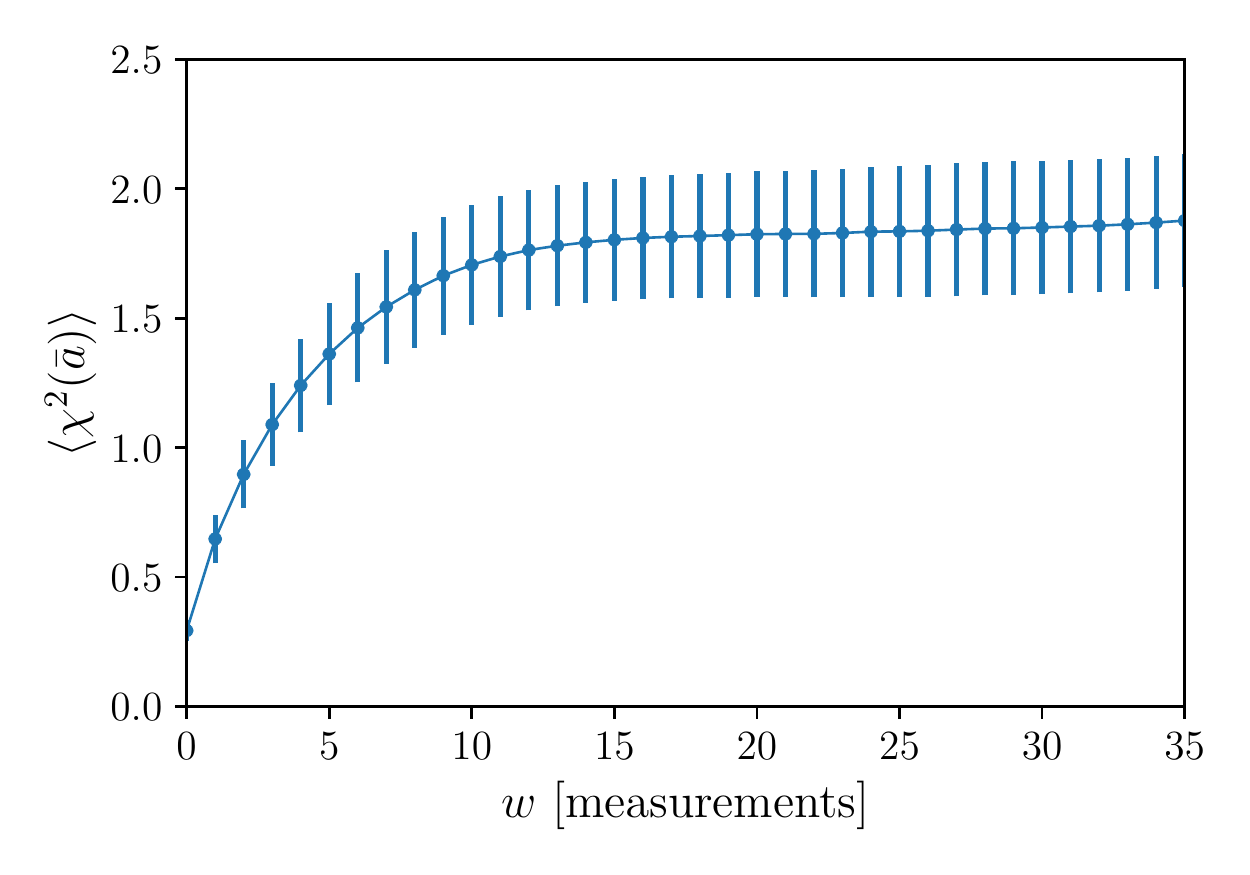}
    \includegraphics[width=.49\textwidth]{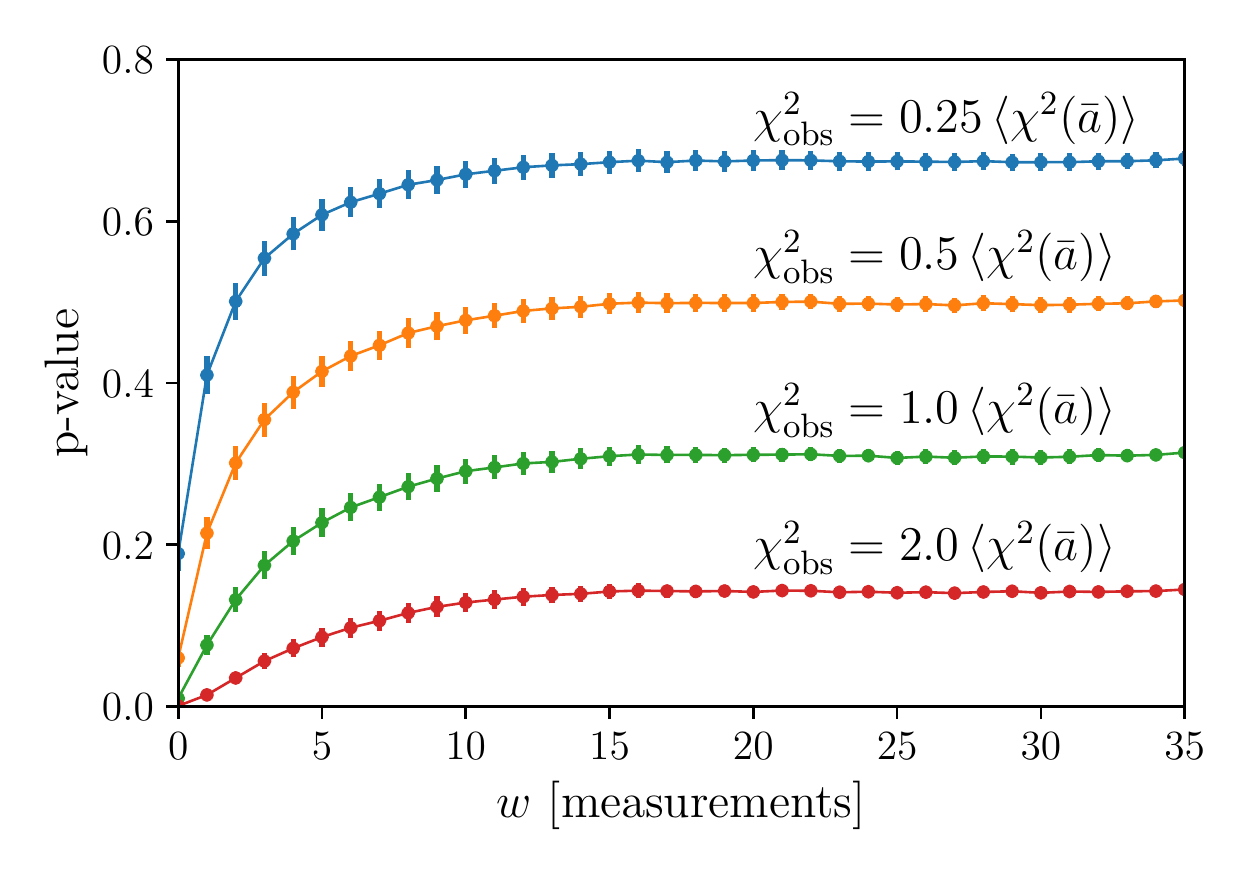}
    \caption{
    Fits of $G_{\phi^2\phi^2}^\mathrm{conn}(t)=G_{\phi^2\phi^2}(t) - \langle \phi^2 \rangle $ in a fit-range $t/a \in [4,26]$.
    Several replica are considered and error bars are obtained from the fluctuations
    over the replica. 
    The left and right panels show the estimator $E_f(w)$, eq.~\eqref{e:res1}, and the p-value 
    for several choices of $\chi^2_\mathrm{obs}$ as a function of the window size $w$ in units of measurements.
    }
    \label{f:autocorr}
\end{figure}
For this setup auto-correlations (cf. the upper panel of \fig{f:autocorrphiphi})
are under control but not negligible, thus mimicking a common situation in lattice QCD. 

We now investigate the determination
of the lowest energy level by a fit to the correlation function 
$G_{\phi^2\phi^2}^\mathrm{conn}(t)=G_{\phi^2\phi^2}(t) - \langle \phi^2\rangle^2 $. It is asymptotically dominated by the (2-particle state, each with momentum $\vec p =\vec{0}$) level 
$a E_\mathrm{exact} = 2 a\omega(\vec{0}) = 2 \cosh^{-1}(1 + a^2 m^2/2) \approx 0.15 $. The higher levels are two-particle states with non-zero momentum of $|\vec p| \geq 2\pi /L $. This leads to a gap of $a \Delta E \approx 0.2$. 
We study the typical approach to the problem of extracting the lowest level through fits to the functional form $A_\mathrm{fit} e^{-m_\mathrm{fit} t}$ to 
 $G_{\phi^2\phi^2}^\mathrm{conn}(t)$.
Of course these suffer from the contaminations $\rmO(\exp(-t\Delta E))$  which are neglected in the fit function. Furthermore the signal-to-noise problem is quite severe for $G_{\phi^2\phi^2}^\mathrm{conn}(t)$. We therefore consider fits only up to an upper cut, which we set to $t_\mathrm{max}/a=26$.

We demonstrate the determination of $\langle \chi^2(\abar)\rangle$ in \fig{f:autocorr} 
for a typical uncorrelated fit. 
Neglecting auto-correlations ($w=0$) leads to a significant  underestimate of 
$\langle \chi^2(\abar)\rangle$. 
However, this is no issue, because one can easily sum up to an appropriate window size $w$, 
even with only 2000 measurements as used here. In fact, the window summation of 
$\langle\chi^2(\abar)\rangle$ saturates earlier than the one of typical observables, 
e.g. $\langle\phi^2\rangle$, \fig{f:autocorrphiphi}. 

\subsection{p-value}

We can identify statistically good fits as those which have an acceptable p-value, say $Q \gtrsim 0.05$. For a correlated fit 
with a known covariance matrix we have $\nu=\P$, and thus 
$\langle \chi^2(\abar)\rangle =\Nx-\NA$. 
Furthermore, its p-value is simply given by
\begin{equation}
    Q(\chi_\mathrm{obs}^2, \P) = 
    \gamma \big( (\Nx-\NA)/2, \chi^2_\mathrm{obs}/2 \big)
    \label{e:pvalue_corr}
\end{equation}
in terms of the upper incomplete $\gamma$-function. 
\begin{figure}[ht]
    \centering
    \includegraphics[width=.49\textwidth]{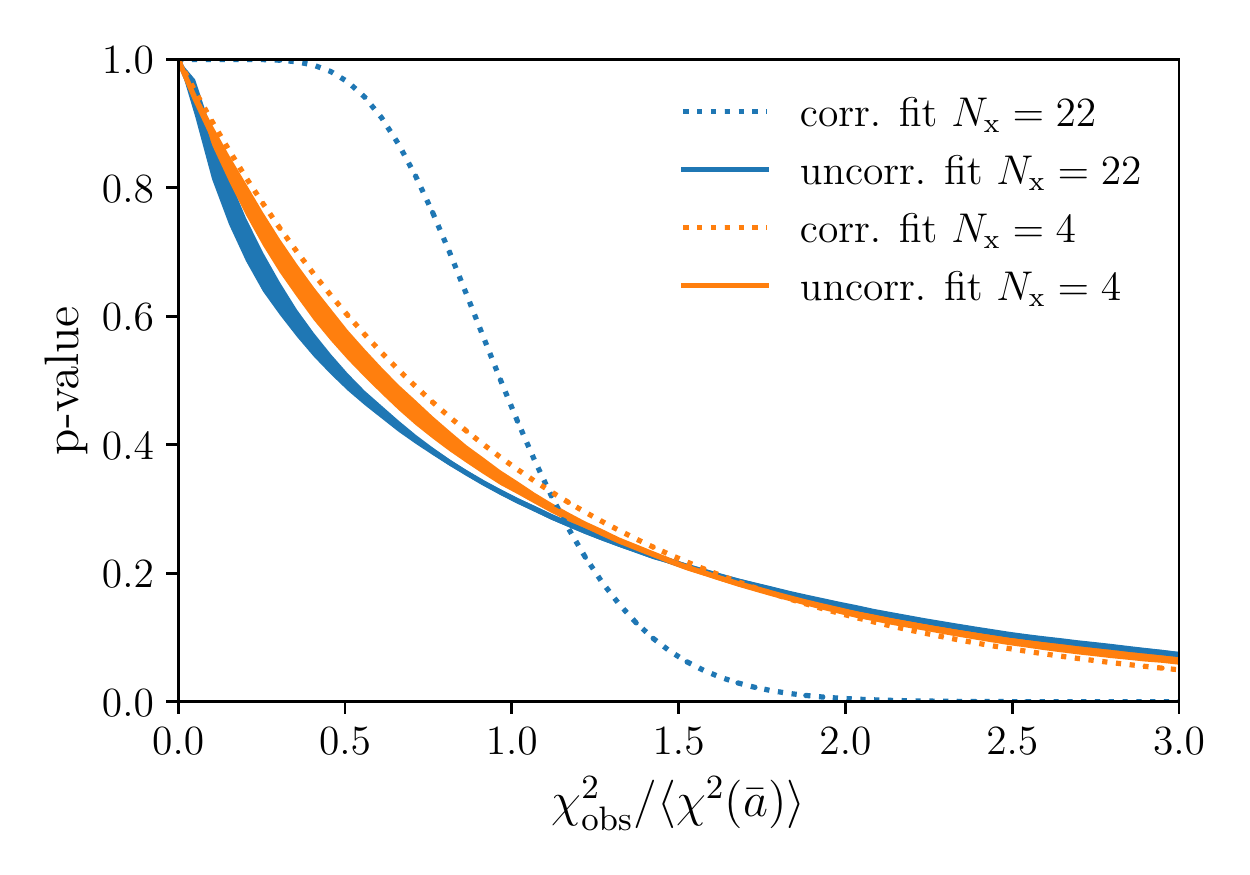}
    \includegraphics[width=.49\textwidth]{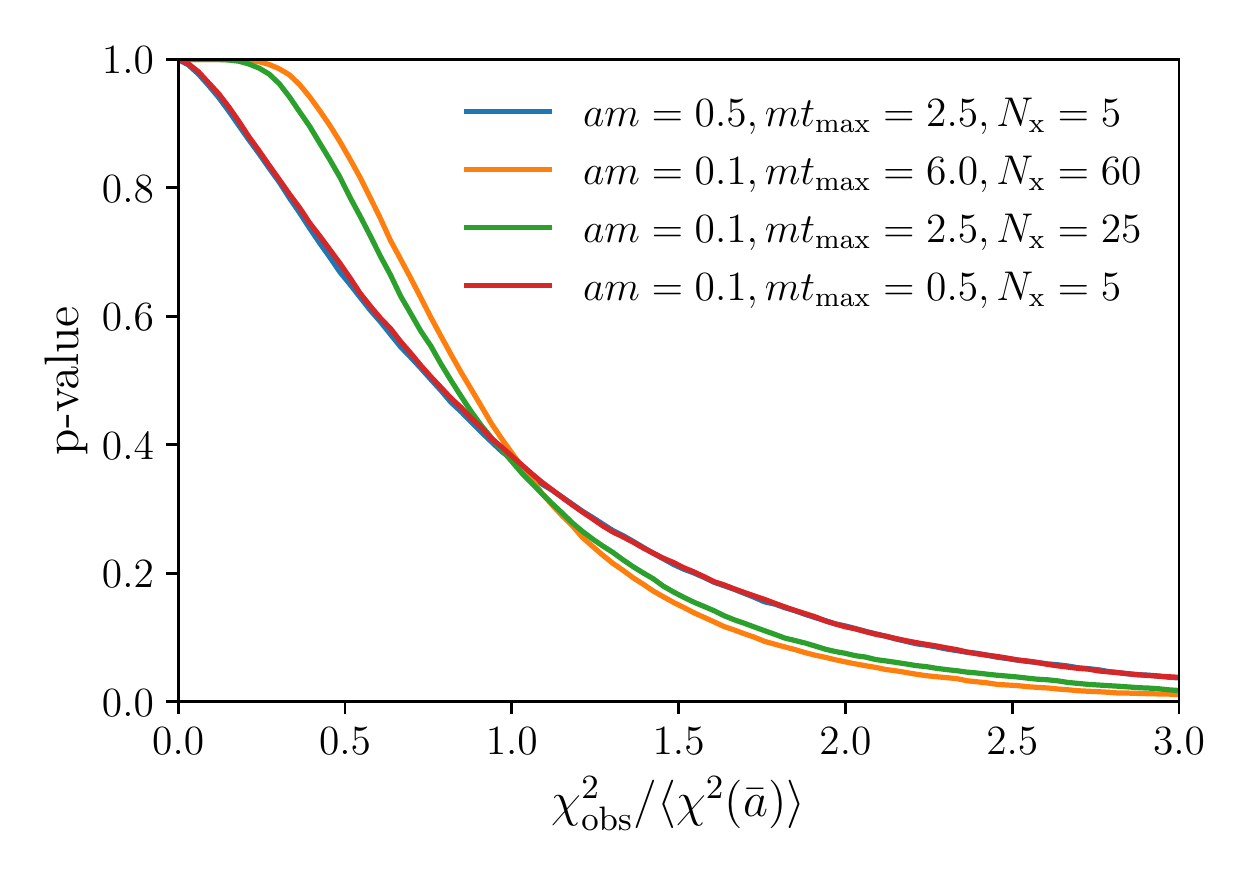}
    \caption{The p-value as a function of the observed $\chi^2$, $\chi^2_\mathrm{obs}$, 
    divided by $\langle \chi^2(\abar) \rangle$. 
    \newline {\em Left: } Fits of $G_{\phi^2\phi^2}^\mathrm{conn}(t)$ are done in a range $t\in[t_\mathrm{min},t_\mathrm{max}]$ where $t_\mathrm{min}/a$ of 4 and 22 are considered, 
    with $t_\mathrm{max}/a=26$.  For correlated fits the exact formula \eq{e:pvalue_corr} is
    illustrated by dashed lines.     For uncorrelated fits we evaluated \eq{e:Qint}.  The statistical error of the p-value curves for uncorrelated
    fits, estimated from several replica, is given by the widths of the curves. 
    \newline {\em Right: } The p-value of uncorrelated fits to $G_{\phi\phi}^\mathrm{conn}(t)$ is evaluated from the exact covariance matrix \eq{e:Gphiphi-cov}. }
    \label{f:pvalue}
\end{figure}
In Fig.~\ref{f:pvalue} we compare this particular case to $Q(\chi^2_\mathrm{obs}, \nu)$, 
\eq{e:Qint}, for uncorrelated fits. Several replica of 2000 measurements are used. For each replicum 
the computation of 
$\langle \chi^2(\abar)\rangle$ employs
the automatic windowing procedure \cite{Wolff:2003sm} for $E_f$. 
The determined window $w_\mathrm{opt}$ is used to
estimate the matrix $\nu$ 
from the covariance matrix. The uncertainty of 
$Q$ due to the statistical fluctuations of $\nu$
is then estimated from the width of the  replica-distribution. 
This very small uncertainty is given by
the widths of the solid lines in the left panel of \fig{f:pvalue}.

While in correlated fits with a {\em large number of degrees of freedom}, a criterion
$\chi^2/\langle \chi^2(\abar) \rangle \approx 1$ is a good discriminator 
between good and bad fits, 
in general the p-value needs to be considered.
Indeed, for uncorrelated fits, it is even more 
important to base the acceptance on the p-value as demonstrated by Fig.~\ref{f:pvalue}. Acceptable p-values are still present for $\chi^2_\mathrm{obs}$ significantly above $\langle\chi^2(\abar) \rangle$ -- also for large $\Nx-\NA$.
The right panel of the figure shows the Q-function 
for a different case, namely a fit to $G_{\phi\phi}$, where we use the analytic covariance matrix without uncertainties. Again values of $\chi^2_\mathrm{obs}/\langle\chi^2(\abar) \rangle$ above 1.5 still have rather good p-values. 
Since $Q$ is easily evaluated, there is no  drawback in always basing the acceptance of a fit on it. Also auto-correlations are easy to control: 
\fig{f:autocorr} shows the dependence of $Q$ on the summation window 
of the auto-correlation functions. One may again just use the 
$w_\mathrm{opt}$ determined in the analysis of $\langle \chi^2(\abar)\rangle$.

\subsection{Determination of the fit window}
At this point we turn to the typical problem of finding the
minimal value of the lower end of the fit range, $t_\mathrm{min}$, 
for which the single exponential
ansatz describes well the data. 
We want to compare the use of the uncorrelated fits
to correlated ones, but the covariance matrices estimated with the statistics of our test-runs 
are (typically) not positive. 
Therefore we generated a long MC chain with 
40000 measurements solely to estimate $C$, and use $W=C^{-1/2}$ in the definition of the correlated $\chi^2$. Of course, this is not a
viable procedure in general  while uncorrelated fits are always possible. 

\begin{figure}[ht]
    \centering
    \includegraphics[width=\textwidth]{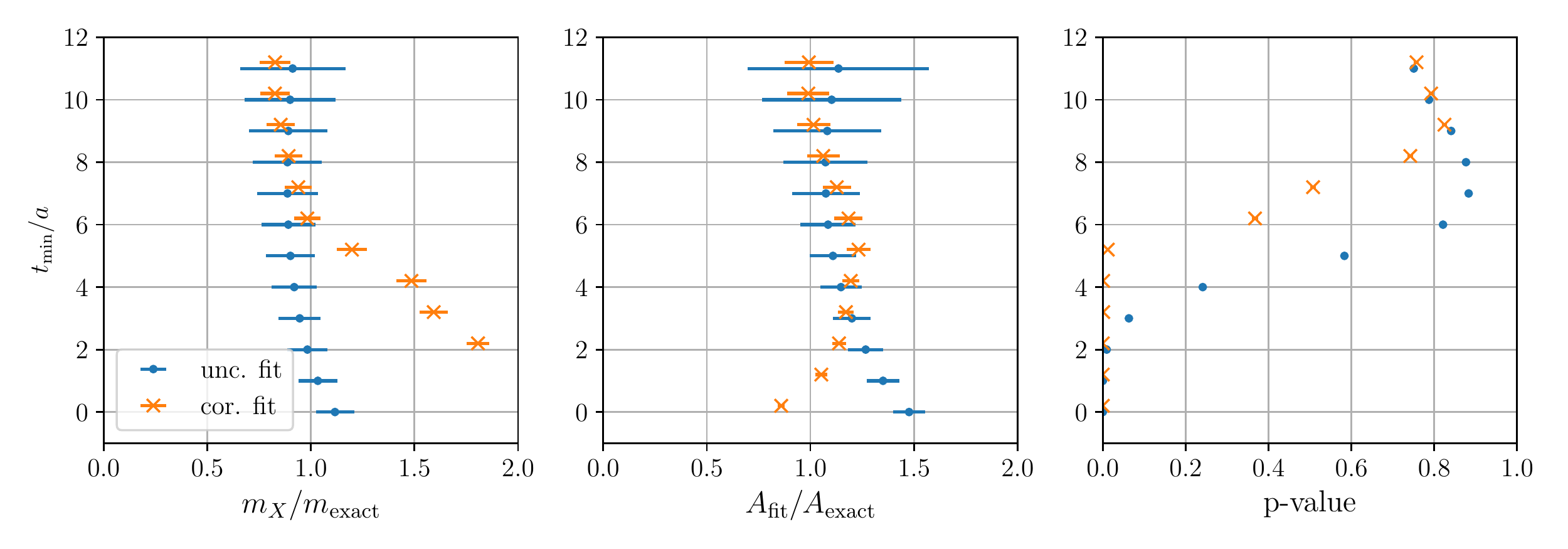}
    \caption{In the left and middle panels we show the fitted masses and amplitudes 
    normalized by their exact asymptotic value as a function of $t_\mathrm{min}$ with $t_\mathrm{max}/a= t_\mathrm{min}/a + 18$. 
    Results for correlated and uncorrelated fits 
    are  represented by orange crosses and blue dots respectively.
    In the right panel we plot the p-values. 
    }
    \label{f:fit_results}
\end{figure}

In Fig.~\ref{f:fit_results} we show one out of  several replica of 4000 measurements each. We compare the results of fits to $G_{\phi^2 \phi^2}^\mathrm{conn}(t)$
to the true asymptotic values of the parameters.
When the p-values becomes acceptable the fitted parameters 
agree reasonably well with their exact asymptotic values.\footnote{We repeated the exercise for larger $L$ with all other parameters the same. In that case the gap, $\Delta E$, is smaller and the good fits need larger $t_\mathrm{min}$. The parameters, in particular the amplitude $A_\mathrm{fit}$, are then  further off when the p-value is not satisfactory.
}
This illustrates that uncorrelated fits can be judged by the computed p-value.  In contrast, the naive goodness of fit criterion, $\langle\chi^2(\abar)\rangle / (\Nx-\NA) \lesssim 1$ is useless, since $\langle\chi^2(\abar)\rangle < 1$ for all fits shown.
We remind the reader that the correlated fits could be done only by estimating the covariance matrix with a 10-folded statistics.

Quite often the data fluctuate more and in a correlated fashion around the true correlation function. One can then have fits with acceptable p-values,
i.e. the fits describe the data over a certain range of a kinematical parameter (here $t$), 
but this does not guarantee that the asymptotic values of the model parameters (here $A$ and $E_\mathrm{exact}$) are determined. 
In our numerical experiment we have seen that in six out of ten cases $A_\mathrm{fit}$ differed 
from the exact value by 30\% to 50\% and more than two standard deviations, even when the p-value was
acceptable. In that sense, the shown replicum is a lucky case.

While we observe smaller statistical errors in the parameters from 
correlated fits for a fixed set of data points,  
uncorrelated fits tend to give
acceptable p-values for smaller $t_\mathrm{min}$.  No particular difference is  then observed in the 
precision of the estimated best fit parameters.

\section{Conclusions}

Estimating the covariance matrix $C$ from limited correlated and auto-correlated data,
in particular the lower part of its spectrum, often proves to be challenging. Here we investigated a method that bypasses
the need for the inverse
covariance matrix and is numerically robust 
against fluctuations of its small eigenvalues.

By studying the expansion of the $\chi^2$ function about its minimum, we have
obtained expressions to compute the expected value of $\chi^2$
and the corresponding p-value, for arbitrary positive weight matrices $W$. The formulae contain the covariance matrix in such a way that 
the upper part of the spectrum of $C$ predominantly determines the expected $\chi^2$ and the p-value. They therefore provide {\em robust} statistical tests for the fits. 
 
We observe that it is important to use the p-value 
to discriminate between good and bad fits, not
just the reduced $\chi^2$.  
An investigation in a toy model illustrates how
the method works with moderate statistics and with auto-correlations present. Unfortunately, 
the p-values of uncorrelated fits are somewhat less sharply discriminating than the ones of correlated fits. This is the price that one has 
to pay for their far superior robustness against
 statistical fluctuations. Thus, when the correlation of the data is well known (very long MC chains),
correlated fits are the preferred choice. Of course, one can also balance the pros and cons of
the two extreme choices by using intermediate weight matrices $W$ together with the proper p-value.

\begin{acknowledgement} 
RS would like to thank U. Wolff and H. Meyer
for sharing their notes on correlated fits. 
RS also recalls B. Bunk mentioning  in the 1980's that
one should estimate the quality of fit for uncorrelated fits from the data.
We would like to thank the members of the ALPHA collaboration for useful discussions.
In particular we  thank A. Ramos for sharing important insights,  
S. Lottini for collaborating at early stages of the project.
MB thanks M.~Hansen, C.~Kelly, N.~Christ, T.~Izubuchi and
C.~Lehner
for several useful discussions on the topic.
We would also like to thank J.~Frison, S.~Kuberski and T.~Vladikas 
for  comments on  earlier versions of 
the manuscript and the referee for constructive comments.
The research of MB is funded through the MUR program 
for young researchers ``Rita Levi Montalcini''.
\end{acknowledgement}

\appendix

\section{Statistical error of $\P$}
\label{a:proof}

Here we derive the explicit form of the projector and other useful
relations. We begin from the minimum condition, which we rewrite as
\begin{equation}
    \P W \dy = W \phi^\alpha(\abar) \da^\alpha + \rmO(N^{-1}) \,.
    \label{e:minimum_P}
\end{equation}
By multiplying on the left by $\phi^{\beta \t}(\abar) W$, we obtain
\begin{equation}
    \da^\alpha = 
    [H^{-1}]^{\alpha\beta} (W \phi^\beta, W \dy) \,, \quad
    H^{\alpha\beta} = (W \phi^\alpha(\abar), W \phi^\beta(\abar)) \,,
    \label{e:da_dy}
\end{equation}
which tells how to propagate the errors from the data to the fitted parameters.
More specifically the covariance matrix of $\abar^\alpha$ reads
\begin{equation}
    C^{\alpha\beta} = [H^{-1}]^{\alpha\alpha'} \, (W \phi^{\alpha'}, \Cw W \phi^{\beta'}) \, 
    [H^{-1}]^{\beta'\beta}  \,.
    \label{e:cov_abar}
\end{equation}

Now combining together \eq{e:da_dy} and \eq{e:minimum_P} one quickly obtains
\begin{equation}
    \P = W \phi^\alpha(\abar) \, [H^{-1}]^{\alpha\beta} \, (\phi^\beta)^\t(\abar) W
\end{equation}
which satisfies all the properties outlined in the main text. 
The partial derivative
\begin{equation}
    \frac{\partial \langle \chi^2(\abar) \rangle}{\partial \abar^\alpha}
    = 2 \tr \big[ \Cw \mathcal K^\alpha (1 - \P) \big] \,,
\end{equation}
combined with \eq{e:cov_abar} gives us the explicit form of the contribution 
to the error of $\langle \chi^2(\abar) \rangle$ that originates from the 
dependence of $\P$ upon $\abar^\alpha$, which we neglect. In the derivation of the result above, 
we have used
\begin{equation}
    \P^\alpha = \frac{\partial \P}{\partial \abar^\alpha} = \mathcal K^\alpha (1-\P) + 
    (1-\P) \mathcal (K^\alpha)^\t \,, \quad
    \mathcal{K}^\alpha = W \phi^{\alpha'}(\abar) \, [H^{-1}]^{\alpha'\beta'} \, (\phi^{\beta'\alpha})^\t(\abar) W \,.
\end{equation}

\section{Reference implementations}
\label{a:impl}

A reference implementation in MATLAB/Octave and Python is publicly available
at this link \url{https://github.com/mbruno46/chiexp}.
Documentation on the syntax can be consulted online,
\url{https://mbruno46.github.io/chiexp}, from
the git repository, locally after downloading it or by
using the corresponding \texttt{help} functions in
the MATLAB or Python interactive sessions.

\section{Generalizations}
\label{a:generalizations}

In the main text we have considered the
case of data described by a single fit function,
but more complicated cases can be easily
accommodated.

When different
sets of data $Y^{(1)}, Y^{(2)}$ (with different sizes $M_1$ and $M_2$)
are described by different
models $\phi^{(1)},\phi^{(2)}$ with a few common parameters,
it suffices to extend $\Nx=M_1+M_2$ to incorporate all
points, such that
\begin{equation}
\begin{split}
    \ybar_i & =  \{ \ybar^{(1)}_1, \dots, \ybar^{(1)}_{M_1}, \ybar^{(2)}_1, \dots ,
    \ybar^{(2)}_{M_2} \} \,, \\
    \phi_i(a) & =  \{ \phi^{(1)}_1(a), \dots, \phi^{(1)}_{M_1}(a),
    \phi^{(2)}_1(a), \dots , \phi^{(2)}_{M_2}(a) \} \,.
\end{split}
\end{equation}
If the two data sets are correlated, e.g. they originate
from the same MC ensembles, but estimating the full $\Nx \times \Nx$
covariance matrix and its inverse turns out to be impractical, the method proposed in this
manuscript allows to arbitrarily regularize the covariance matrix
in the definition of $\chi^2$ (e.g. just consider the diagonal blocks
$M_1 \times M_1$ and $M_2\times M_2$) and still have a reliable
statistical interpretation of the associated fit.

Similarly, there are situations where
we might want to include the error of the kinematic
coordinates $\bar x_i$ (for instance they might be
obtained from averages over MC chains which may also be correlated
with the data points $\ybar_i$). Such cases are easily incorporated 
by considering the vectors
\begin{equation}
    r_i = \{ \ybar_1 - \Phi(X_1,A),\dots,\ybar_{\Nx} - \Phi(X_{\Nx},A),
    \bar x_1 - X_1, \dots \bar x_{\Nx} - X_{\Nx} \}
\end{equation}
in the definition of $\chi^2 = || W r||^2$.
The matrix $W^2$ may correspond to various regularizations of
the inverse of the extended covariance matrix
\begin{equation}
    C = \Big\langle \, \begin{matrix} \dy \dy^\t & \dy \delta \bar x^\t \\
    \delta \bar x \dy^\t & \delta \bar x \delta \bar x^\t
    \end{matrix} \, \Big\rangle\,,
\end{equation}
and nothing changes in the derivation of $\langle \chi^2(\abar)\rangle$
and in the discussion on the goodness-of-fit
for the model function $\Phi$.
In the equations above $X_i$ represent the true values of
the kinematic coordinates, and $\delta \bar x_i = \bar x_i - X_i$.

\section{Alternative estimators of the covariance matrix}
\label{a:errors}

In this Appendix we discuss the replacement of the $\Gamma$-method by
other error estimation techniques.

\subsection{Jackknife and binning methods}
 
When it is known that the (exponential) auto-correlation time
is small compared to the run length $N_\ens$, the covariance matrix
may be estimated by Jackknife resampling
of binned data, with bin length significantly
larger than the auto-correlation time.
Assuming that $N_\ens = B_\ens N_B^\ens$, blocked measurements
on a given ensemble are defined as
\begin{equation}
    b_i^\ens(k) = \frac{1}{B_\ens} \sum_{t=1}^{B_\ens} p_i^\ens((k-1) B_\ens + t) \,, \quad
    k = 1, \dots , N_B^\ens \,,
\end{equation}
while complementary or jackknife bins are 
\begin{equation}
    c_i^\ens(k) = \frac{N_B^\ens \pbar_i^\ens - b_i^\ens(k)}{N_B^\ens - 1} \,.    \label{e:jackknife}
\end{equation}
An estimator of the covariance matrix is given by
\begin{equation}
    C_{ij} = \sum_{\ens} \frac{\NB^\ens-1}{\NB^\ens}
    \sum_{k=1}^{\NB^\ens} \Delta c^\ens_i(k) \Delta c^\ens_j(k) \,,
    \quad \Delta c_i^\ens(k) = c_i^\ens(k) - \pbar_i^\ens \,, 
    \label{e:cov_jack}
\end{equation}
and the case of derived observables $Y_i=\eta_i(P)$
is obtained by replacing $c_i^\ens(k)$ with $\eta_i(c^\ens(k))$ in the equations above.
The generalization of our previous results to jackknife estimators is
straightforward: starting from the intermediate quantities
\begin{eqnarray}
    E_f^\ens (k) &=& [\Delta c^\ens(k)]^\t W(1 - \P) W
    \Delta c^\ens(k) \,, \\
    E_s^\ens (k) &=& [\Delta c^\ens(k)]^\t W\P W
    \Delta c^\ens(k) \,,
\end{eqnarray}
the central value of $\langle \chi^2(\abar) \rangle$ is obtained by
replacing $E_{f,s}$ in \eq{e:chiexp-ef} and \eq{e:chiexp-es} with
\begin{equation}
    E_{f,s} = \sum_\ens \frac{N_B^\ens-1}{N_B^\ens}
    \sum_{k=1}^{N_B^\ens} E_{f,s}^\ens(k) \,.
\end{equation}
Using the results of Ref.~\cite{Wolff:2003sm}
one can estimate their errors according to
\begin{equation}
    [\Delta E_{f,s}]^2 \approx 
        2 \sum_\ens \frac{1}{N_B^\ens}  \big[E_{f,s}^\ens\big]^2 \,,
\end{equation}
whereas the p-value is estimated from the covariance matrix
in \eq{e:cov_jack}.

\subsection{Priors and the Bayesian approach}

Gaussian priors on the parameters, a special case of a Bayesian analysis,
are easily taken into account by extending the $r$ vectors for $\chi^2 = ||W r||^2$ to
\begin{equation}
    r_i = \{ \ybar_1 - \Phi(x_1,a),\dots,\ybar_{\Nx} - \Phi(x_{\Nx},a),
    a_1 - \widetilde a_1, \dots , a_{N_\mathrm{priors}} - \widetilde a_{N_\mathrm{priors}} \}
\end{equation}
and the covariance matrix to
\begin{equation}
    C = \Big\langle \, \begin{matrix} \dy \dy^\t & 0 \\
        0 & \mathrm{diag}[1/\sigma_{\widetilde a_1}^2, \dots, 1/\sigma_{\widetilde a_{N_\mathrm{priors}}}^2]
    \end{matrix} \, \Big\rangle\,,
\end{equation}
with $\widetilde a^\alpha$ and $\sigma_{\widetilde a^\alpha}$ representing the prior and its 
width for the parameter $a^\alpha$.
One can easily verify that for correlated fits $\langle \chi^2(\abar)\rangle = \Nx + N_\mathrm{priors}-\NA
$. 

\subsection{Master fields}
\label{s:master}

Recently M.~L\"uscher has derived formulae  \cite{Luscher:2017cjh} for 
statistical errors based on the invariance under translations of space-time, rather than of Monte-Carlo time. We write  
\begin{equation}
    \ybar_i = \frac{1}{V} \sum_s y_i(s) \,, \quad 
    \lim_{V \to \infty} \ybar_i = Y_i\,,
\end{equation}
with $s$ labelling the sites for a given large-volume configuration,
called master field. Following Ref.~\cite{Albandea:2021lvl},
the $\Gamma$ method is then defined from the counter-part of 
$\Gamma^{\MC}(t)$, which we may now call $\Gamma^\mathrm{MF}_{ij}(s)$,
and which is evaluated from the fluctuations 
$\dy_i (s) = y_i(s) - \ybar_i$. As described in Ref.~\cite{Luscher:2017cjh},
the variance of $\ybar_i$ may be obtained by summing $\Gamma^\mathrm{MF}_{ii}(s)$ 
up to distances $|s|<R$, thus introducing only an exponentially small bias 
of order $e^{-2m_\pi R}$.
Our formulae then hold after 
replacing $\Gamma^{\MC}_{ij}$ with $\Gamma^\mathrm{MF}_{ij}$ 
e.g. in \eq{e:chiexpac} and
after providing an appropriate windowing procedure.

\clearpage
\usebiblio{biblio}

\end{document}